# Deep interval prediction model with gradient descend optimization method for short-term wind power prediction

Chaoshun Li[*,1], Geng Tang[1], Xiaoming Xue[*,2], Xinbiao Chen[1], Ruoheng Wang[1], Chu Zhang[2]

[1]*School of Hydropower and Information Engineering, Huazhong University of Science and Technology, Wuhan 430074, China;*
[2] *Faculty of Mechanical and Material Engineering, Huaiyin Institute of Technology, Huai'an 223003, China*

**Abstract:**

The application of wind power interval prediction for power systems attempts to give more comprehensive support to dispatchers and operators of the grid. Lower upper bound estimation (LUBE) method is widely applied in interval prediction. However, the existing LUBE approaches are trained by meta-heuristic optimization, which is either time-consuming or show poor effect when the LUBE model is complex. In this paper, a deep interval prediction method is designed in the framework of LUBE and an efficient gradient descend (GD) training approach is proposed to train the LUBE model. In this method, the long short-term memory is selected as a representative to show the modelling approach. The architecture of the proposed model consists of three parts, namely the long short-term memory module, the fully connected layers and the rank ordered module. Two loss functions are specially designed for implementing the GD training method based on the root mean square back propagation algorithm. To verify the performance of the proposed model, conventional LUBE models, as well as popular statistic interval prediction models are compared in numerical experiments. The results show that the proposed approach performs best in terms of effectiveness and efficiency with average 45% promotion in quality of prediction interval and 66% reduction of time consumptions compared to traditional LUBE models.

| Nomenclature | | | |
|---|---|---|---|
| $W_{fh}$ | weights matric of forget gate for connection of last cell output | $L_i$ | lower bound using *i* th sample. |
| $W_{fx}$ | weights vector of forget gate for connection of current input | $y_i$ | observe value of *i* th sample. |
| $W_{ih}$ | weights matric of input gate for connection of last cell output | $W_v$, $b_v$ | weights and bias matrices in fully connected layers. |
| $W_{ix}$ | weights vector of input gate for connection of current input | $(W, b)$ | all weights and bias matrices in deep leaning network |

[*] Corresponding author, Chaoshun Li and Xiaoming Xue
Email: csli@hust.edu.cn; taxueli@hyit.edu.cn

| | | | |
|---|---|---|---|
| $W_{oh}$, | weights matric of output gate for connection of last cell output | $f_1$, $f_2$ | two target functions |
| $W_{ox}$ | weights vector of output gate for connection of current input | $d$ | distance between $y_i$ and boundary of $u_i$ and $l_i$. |
| $W_{ch}$ | weights matric of input function for connection of last cell output | $\lambda$ | penalty coefficient of target function $f_1$. |
| $W_{cx}$ | weights vector of input function for connection of current input | $g^1$, $g^2$ | average gradients of $f_1$ and $f_2$ |
| $b_f$ | bias vector of forget gate | m | the number of samples used to calculate the average gradient. |
| $b_i$ | bias vector of input gate | $r_t$ | the adjustment parameter of learning rate at update time $t$ |
| $b_o$ | bias vector of output gate | $\rho$ | decay rate. |
| $b_c$ | input bias vector | $w$ | abbreviation of ($W$, $b$) |
| $C_r$ | State matric of $r$th LSTM cell | $\delta$ | a small value used to avoid division by zero in the calculation of learning rate |
| $H_r$ | Output matric of rth LSTM cell | n | the number of testing samples |
| $f_r$ | output matrices of input gate | A | the range of the target variable |
| $o_r$ | output matrices of output gate | μ | expected value of PICP |
| $\tilde{c}_r$ | candidate value of cell state | η | penalty parameter of PICP |
| $R$ | number of LSTM cells | α, β | hyper-parameters in $CWC_{proposed}$ |
| $x_i$ | $i$ th sample. | σ | width factor |
| $u_i$, $l_i$ | output of fully connected layer using $i$ th sample. | Rand | a random number between 0 and 1 |
| $U_i$ | upper bound using $i$ th sample. | | |

## 1. Introduction

With the gradual depletion of traditional fossil energy and the increasingly serious air pollution, renewable energy has attracted more and more attention. Wind power is becoming increasing popular because of its clean and recyclable nature. However, due to the nonlinear and nonstationary characteristics of the wind power, it brings severe challenges to the safety and reliability of power systems. Therefore, high quality wind power prediction is of great significance and practicability for making an optimal power system planning, reasonably arranging system reserve.

There are two basic categories of wind power prediction methodologies namely point prediction [1][21][17][5][4] and interval prediction[32][15][45]. High prediction accuracy and the elimination of prediction error are the everlasting goals for point prediction methods. However, in real word applications, prediction error is impossible to obliterate and the uncertainty of prediction must also be quantified. Compared with point prediction, the interval prediction provides prediction intervals (PIs), which directly communicate uncertainty, offering a lower and upper bound with the assurance that the estimated data will fall between the bounds. As for wind power interval prediction (WPIP), PIs of wind power instead of accurate values will give a more comprehensive reference to the planning and operation of power systems.

Many methods have been suggested by the researchers with a prior assumption for probabilistic forecasting to construct the prediction intervals. Traditionally Bayesian [11], mean-variance [25] and Bootstrap [2] methods are used for the construction of prediction interval. However, these methods exist some drawbacks. The Bayesian method has high computational cost and imprecise PIs. The mean-variance estimation method suffers from low empirical coverage probability. Although Bootstrap method is simple and easy to implement, main problem associated with this method is time consuming. Some statistical methods without any prior assumption, like the quantile regression method [42], the kernel density forecast method [1], and the Gaussian process [41], may suffer from restrictive assumptions about data distribution and depend on the quantile analysis results of point prediction.

To overcome problems of statistical model, Khosravi proposed a more reliable approach called Lower upper bound estimation (LUBE) method [12], in which an interval prediction model constructed by a neural network with two outputs for estimating the prediction interval bounds was designed. The cost function of coverage width-based criterion (CWC) consisting of two indices, namely the interval width and the interval coverage probability, is used as the objective function for model optimization and meta-heuristic optimization algorithms [10][38][18] could be applied to tune the interval prediction model. Because of the outstanding design, interval prediction models based on LUBE method have been developed widely in various applications of interval prediction. Typically, two categories of LUBE methodologies have been developed.

(1) Different core models. In the LUBE frame, different types of basic forecasting models could be integrated and the corresponding interval prediction models, like support vector machine (SVM) based and ELM based LUBE methods [31][23][40], have been constructed.

(2) Different optimization methods. Single and multiple objective meta-heuristic optimization (MHO) algorithms have been applied to train the LUBE models. The CWC function integrating indices of interval width and the interval coverage probability, is used as the single objective function for model optimization. The indices of interval width and interval coverage probability can be optimized directly as two separate objectives, and a multi-objective optimized LUBE model [32][43] could then be designed with an effective multi-objective optimization algorithms [13][14][16].

Since various machine learning models, like ANN, SVM and ELM could be implemented in the LUBE framework, deep learning models could also be feasible and even achieve better performance than those sallow, considering the outstanding learning capacity confirmed in numerous applications [36][20]. Deep learning conforms to the trend of big data and has strong learning and generalization ability for massive data [28]. Some popular deep learning methods, such as convolutional neural network (CNN) [35], deep auto-encoder (DAE) [9] and long short-term memory (LSTM) network [27][39][19], have already been introduced in the field of the wind forecasting, due to their advantage over traditional forecasting models. It seems sound that a LUBE method constructed by a deep learning model would be powerful. Before that, the questions will be a little disappointing: how can we build it? Will the traditional design and training methods still work?

Though the LUBE models have gained great popularity, a great limitation could not be neglected. The training method for traditional LUBE models is always based on meta-heuristic optimization, for the gradient descent (GD) method is incompatible with the training process. The traditional CWC function [12] is not continuous and differentiable, thus it could not be used as the loss function to construct GD training method. Although MHO, like Annealing (SA), Particle Swarm Optimisation (PSO), have successfully been applied to design training method, this is inconvenient since GD has become the standard training method for NNs [6]. The complexity and difficulty of applying MHO to train LUBE would even become unacceptable if the number of optimization variables increase to a certain extend. Considering there might be thousands parameters needing to optimize for deep learning models, the MHO method would be extremely time-consuming and difficult to get the feasible solutions, for a possible training of a deep learning model based LUBE method.

Based on the discussion above, this paper try to propose a deep learning model based LUBE method for wind power prediction and resolve the training problem by design a GD method. To our best knowledge, it is the first time that a LUBE framework for WPIP is built with an efficient GD training method.

The contributions of this paper include: (1) A novel deep interval prediction framework is designed and the LSTM is chosen as representative to show the LUBE model is constructed; (2) new loss functions are designed for applying root mean square back propagation (RMSprop) method for efficient GD training; (3) a novel CWC function has been designed for evaluation LUBE method to enhance the performance of prediction. The deep learning model for interval prediction has been applied in wind power prediction and fully tested by comparing with traditional LUBE methods based on KELM, SVM, and ANN in comparative experiments.

The rest of the paper is presented as followings: Section 2 introduces theoretical backgrounds of LSTM network and RMSprop method. Section 3 describes the new approach for the method of wind power interval prediction (WPIP). Section 4 presents the detail of the experiment and results. Finally, Section 5 summarizes the conclusions.

## 2. Theoretical backgrounds of long short-term memory

LSTM network [7] belongs to the family of deep recurrent neural network (RNN). By incorporating self-connected "gates" in the hidden units, the LSTM method tends to solve the vanishing gradient problem, one of the major drawback associated with standard RNNs. An LSTM network is composed of basic units called memory cell. The structure of the LSTM cell is presented in Fig. 1.

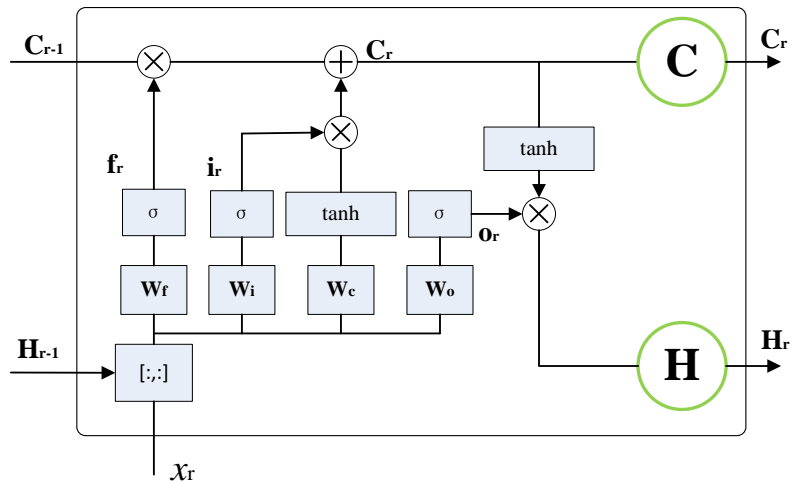

**Fig. 1** Architecture of LSTM cell

An LSTM cell consists of an input gate, forget gate and output gate. The forget gate determines how much of prior memory value should be removed from the cell state. Similarly, the input gate specifies new input to the cell state and output gate control the value of output. The gate is actually a layer of fully connected layers, which is defined as follows:

$$f_r = \sigma(W_{fh}H_{r-1} + W_{fx}x_r + b_f) \tag{1}$$

$$i_r = \sigma(W_{ih}H_{r-1} + W_{ix}x_r + b_i) \tag{2}$$

$$o_r = \sigma(W_{oh}H_{r-1} + W_{ox}x_r + b_o) \tag{3}$$

$$\tilde{C}_r = \tanh(W_{ch}H_{r-1} + W_{cx}x_r + b_c) \tag{4}$$

where $H_{r-1}$ and $x_r$ implies the output of last LSTM and input of current LSTM respectively; $\tilde{c}_r$ is candidate value. The activation function of $\sigma(\cdot)$ is usually the Sigmoid function.

Then, the cell state $C_r$ and output value $H_r$ is calculated as:

$$C_r = f_r \circ C_{r-1} + i_r \circ \tilde{C}_r \tag{5}$$

$$H_r = o_r \circ \tanh(C_r) \tag{6}$$

where ∘ denotes the Hadamard product.

The training methods are decisive in destemming the performance of deep learning networks. Some popular training methods, like batch gradient descent (BGD), resilience propagation and RMSprop [44][3], have been developed. Among these methods, the RMSprop has been proved to be an effective and practical deep learning network optimization algorithm [30][26]. The RMSprop stores a decaying average of past gradients and squared gradients, and introduces the concept of local learning to avoid attenuation of global learning rate and enable fast training and satisfactory convergence. As it has been proved that these improvements are effective in enhancing the learning ability of the LSTM network.

## 3 The interval prediction model based on long short-term memory and Lower upper bound estimation

LSTM has been successfully applied to develop point prediction models due to its advantages over traditional networks. However, the interval prediction model based on LSTM has not been developed to the best of our knowledge. In this section, the interval prediction model based on LSTM and LUBE is designed and illustrated in detail.

### 3.1 Frame of the new model

The interval prediction model employs a deep leaning architecture containing three parts, namely the LSTM network, the fully connected layers and the rank ordered terminal, as shown in Fig. 2 (a). The LSTM network of the first part undertakes the task of transforming the original input $x = (x_1, x_2, \ldots, x_R)$ into high dimensional vector $H_R$. After then, the output of the final LSTM cell $H_R$ is imported to the fully connected layers. In this context, the final LSTM cell possesses the greatest impact on the prediction result and the impact of other cell, as the number descend, decays gradually. This characteristic of model structure suits well to the time series prediction, which means the nearer samples take more weights in determining the forthcoming sample that needs to be predicted.

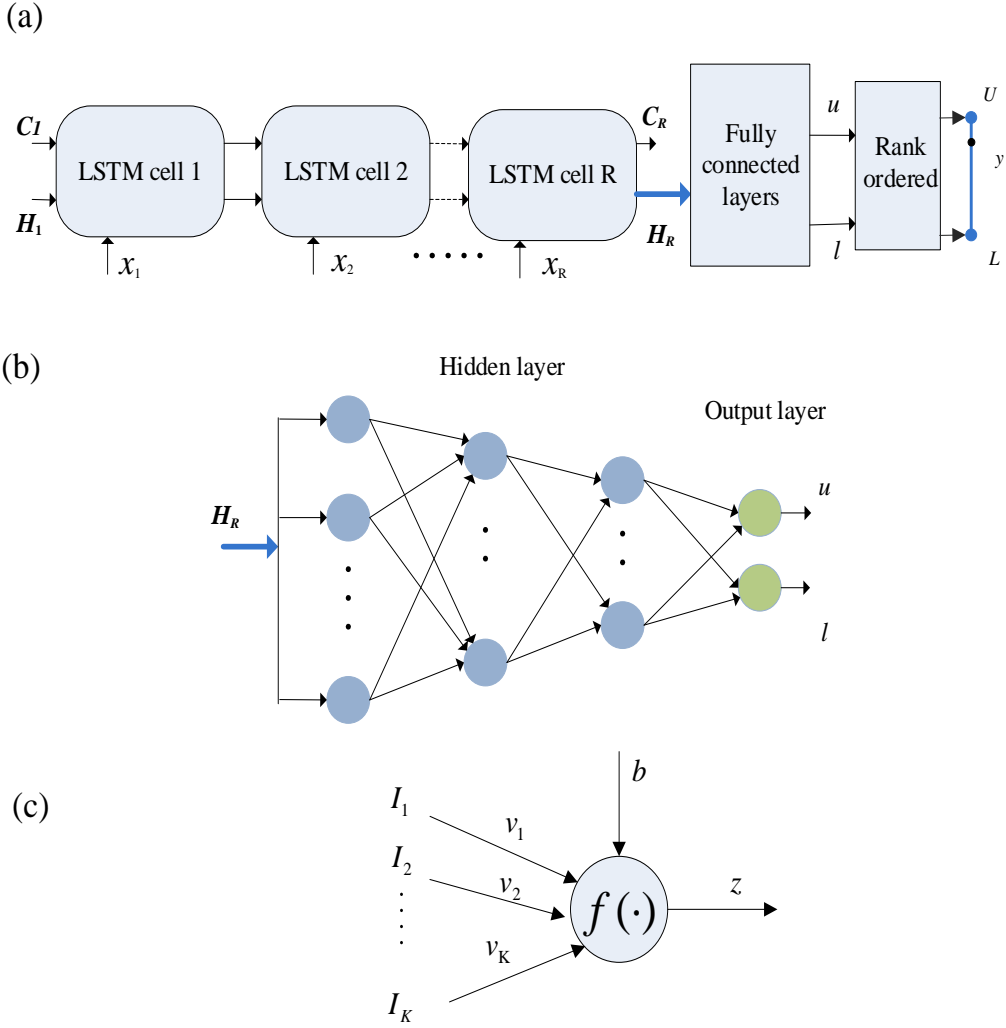

**Fig. 2** Architecture of interval prediction using deep leaning network

The second part of the proposed model consists of fully connected layers, which aims to produce the lower and upper bounds for interval construction. The structure of the fully connected layers is exhibited in Fig. 2(b). The input vector $\boldsymbol{I} = (I_1, I_2, ..., I_K)$ of each neuron consists the outputs of neurons of the connected preceding layer. The parameters to be optimized are $v$ and $b$, where $\boldsymbol{v} = (v_1, v_2, ..., v_K)$ represents the connection weights between this neuron and the neurons in its preceding layer, and b is a threshold value or bias. The output of the neuron $z$ is defined in Eq. (7) and $f$ is activation function. The gray cells are hidden layers whose activation function is the Relu function as defined in Eq. (8). Compared with Sigmoid function, Relu function has the advantages of calculation efficiency, avoiding gradient disappearance and being able to get a lower activation rate. The two green cells are output layers whose activation function is the pure linear function as defined in Eq. (9). The output $z$ of two green cells are also the output fully connected layers, which are defined as $u$ and $l$.

$$z = f(\boldsymbol{I} \cdot \boldsymbol{v} - b) \tag{7}$$

$$f(x) = \max(0, x) \tag{8}$$

$$f(x) = x \tag{9}$$

In the third part, the upper bound and lower bound are constructed by ranking the outputs of $u$ and $l$. The formulations are defined as:

$$U = max(u,l) \tag{10}$$

$$L = min(u,l) \tag{11}$$

where $U$ and $L$ are upper bound and lower bound respectively.

## 3.2 Model training strategy

Different from point prediction, interval prediction model aims to obtain PI with narrow PI width and larger PI coverage probability. To fill these goals, two objective functions are designed for model training. On the one hand, the observed value is expected to be closer to the midpoint of PI, when the observed value is in PI. Once the observed value fall out the PI, the value of objective function should be punished. The farther the observed value departure the PI, the higher the penalty will be. On the other hand, the width of PI is expected to become narrower. Based on these principles, two target function, $f_1$, and $f_2$, are defined in Eq. (12) and Eq. (13) respectively.

$$f_1(\boldsymbol{W},\boldsymbol{b}) = k_1 \cdot \left( \left| y_i - \frac{u_i + l_i}{2} \right| + \lambda \cdot \gamma \cdot d \right) \tag{12}$$

$$f_2(\boldsymbol{W},\boldsymbol{b}) = k_2 \cdot \left| u_i - l_i \right| \tag{13}$$

where $k_1$ and $k_2$, which are the weights of $f_1$ and $f_2$, can determine importance between hit rate and width of PI. $y_i$ is the observed value of $i$ th sample $\boldsymbol{x_i}$. $u_i$ and $l_i$ are defined in Eq. (14), in which $\Psi_u$ and $\Psi_l$ are functions of the relationship among deep leaning network. $(\boldsymbol{W},\boldsymbol{b})$ contains ($\boldsymbol{W_{fh}}$, $\boldsymbol{W_{ih}}$, $\boldsymbol{W_{oh}}$, $\boldsymbol{W_{ch}}$, $\boldsymbol{W_{fx}}$, $\boldsymbol{W_{ix}}$, $\boldsymbol{W_{ox}}$, $\boldsymbol{W_{cx}}$, $\boldsymbol{b_f}$, $\boldsymbol{b_i}$, $\boldsymbol{b_o}$, $\boldsymbol{b_c}$) in LSTM network and weight matrix $\boldsymbol{V}$ in fully connected layers. $\lambda$ is penalty coefficient and $\gamma$ is step function which is defined in Eq. (15). $d$ is the distance between $y_i$ and boundary of $u_i$ and $l_i$, when $y_i$ is out of PI. Finally, d is defined in Eq. (16).

$$\begin{cases} u_i = \psi_u((\boldsymbol{W},\boldsymbol{b}), \boldsymbol{x}_i) \\ l_i = \psi_l((\boldsymbol{W},\boldsymbol{b}), \boldsymbol{x}_i) \end{cases} \tag{14}$$

$$\gamma = \begin{cases} 0, y_i \text{ is between } l_i \text{ and } u_i \\ 1, y_i \text{ is out of } l_i \text{ and } u_i \end{cases} \tag{15}$$

$$d = \left| y_i - \frac{u_i + l_i}{2} \right| - \left| \frac{u_i - l_i}{2} \right| \tag{16}$$

The $(\boldsymbol{W}, \boldsymbol{b})$ represent all the weights and bias in the proposed LSTM model, which are the parameters that need to solve. The model is trained by means of RMSprop algorithm, and the parameters could be updated in training process. The formula of average gradient $[g_1]_t$ and $[g_2]_t$ from objective function $f_1(W, b)$ and $f_2(W, b)$ is given by

$$\begin{cases} [g_1]_t = \dfrac{1}{m} \nabla_w \sum_{j=1}^{m} L(f_1(x_j, w), y_j) \\ [g_2]_t = \dfrac{1}{m} \nabla_w \sum_{j=1}^{m} L(f_2(x_j, w), y_j) \end{cases} \tag{17}$$

$$g_t = [g_1]_t + [g_2]_t \tag{18}$$

where m is the number of samples. For the sake of simplification, $w$ represent $(W, b)$ for convenience. The update function of learning rate $r$ at time $t$ is given by

$$r_t = \rho \cdot r_{t-1} + (1 - \rho) \cdot (g_t)^2 \tag{19}$$

where $\rho$ is the decay rate. The initial value $r_0$ is zero.

The update to the weights is then given by

$$w_t = w_{t-1} - \frac{\varepsilon}{\delta + \sqrt{r_t}} \cdot g_t \tag{20}$$

where ε is the global learning rate. The value δ is a small value used to avoid division by zero.

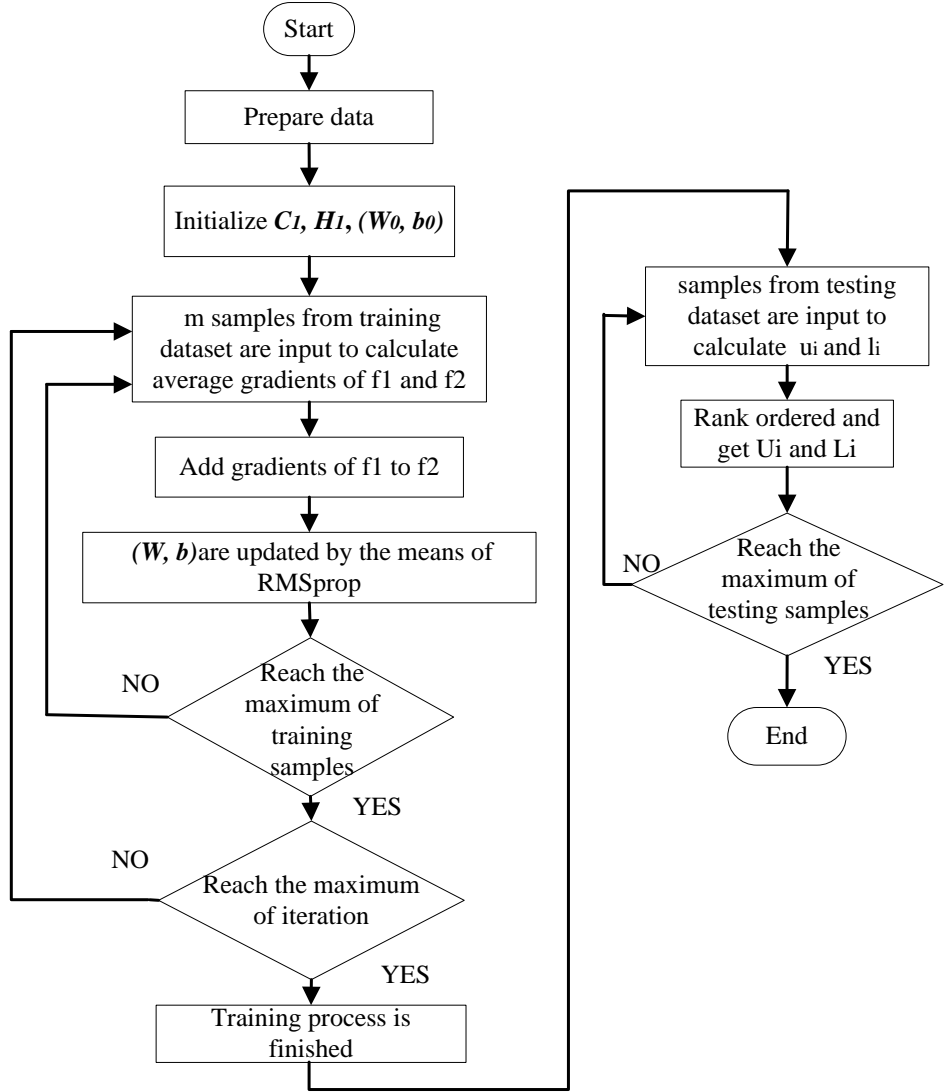

**Fig. 3** The flowchart of interval prediction

### 3.3 Procedures of the proposed approach

The flowchart of the new approach for WPIP is shown in Fig. 3, and the detailed steps are presented as follows.

**Step 1**: Prepare data. Wind power time series is processed to construct the testing dataset and training dataset. Initialize $C_1$, $H_1$, $(W_0, b_0)$.

**Step 2:** Train the deep leaning network.

Step 2.1: m samples from training dataset are input to calculate average gradients of $f_1$ and $f_2$.

Step 2.2: Calculate the sum of gradient from $f_1$ and $f_2$.

Step 2.3: $(W, b)$ are updated by the means of RMSprop.

Step 2.4: Repeat 2.1 to 2.3 until reaching the maximum of training samples.

Step 2.5: Repeat 2.1 to 2.4 until reaching the maximum of iteration.

**Step 3:** Test the deep leaning network.

Step 3.1: samples from testing dataset are input to calculate $u_i$ and $l_i$, which is not ranked.

Step 3.2: Rank $u_i$ and $l_i$. Assign the greater value among $u_i$ and $l_i$ to $U_i$ and the smaller value among $u_i$ and $l_i$ to $L_i$.

Step 3.3: Repeat 3.1 to 3.2 until reaching the maximum of testing samples.

## 3.4 Indices for model evaluation

Prediction Interval Coverage Probability (PICP) [43] is defined as:

$$PICP = \left(\frac{1}{n}\sum_{i=1}^{n} C_i\right) \cdot 100\% \tag{21}$$

where $n$ is the number of testing samples and the expression of $C_i$ is presented as:

$$C_i = \begin{cases} 1, y_i \in [L_i, U_i] \\ 0, y_i \notin [L_i, U_i] \end{cases} \tag{22}$$

where $U_i$ and $L_i$ are the upper and lower bound and $y_i$ is the observed value of i th sample.

Prediction Interval Normalized Average Width (PINAW) and Prediction Interval Normalized Root-mean-square Width (PINRW) [43] defined as:

$$PINRW = \frac{1}{A}\sqrt{\frac{1}{n}\sum_{i=1}^{n}(U_i - L_i)^2} \tag{23}$$

$$PINAW = \frac{1}{nA}\sum_{i=1}^{n}(U_i - L_i) \tag{24}$$

where $U_i$, $L_i$ and n are the same as in PICP and $A$ is the range of the target variable.

The Normalized average deviation (NAD) [30] is used to express the deviation of the data which are not covered by the PI. So it can express the rationality of PI.

$$I_{NAD} = \frac{1}{n}\sum_{i=1}^{n} a_i \tag{25}$$

where the expression of $a_i$ is defined in Eq. (26).

$$a_i = \begin{cases} (L_i - t_i)/\frac{1}{n}\sum_{i=1}^{n}(U_i - L_i), t_i < L_i \\ 0, t_i \in [L_i, U_i] \\ (t_i - U_i)/\frac{1}{n}\sum_{i=1}^{n}(U_i - L_i), t_i > U_i \end{cases} \tag{26}$$

where $U_i$ and $L_i$ and n are the same as in PICP and $t_i$ is the observed value of ith sample.

In order to account for both the width as well as coverage probability in a comprehensive manner, a comprehensive index consisting of both PICP and PINAW, known as coverage width criterion (CWC) [32] was developed, which is defined as:

$$CWC_{original} = \begin{cases} PINAW, PICP >= \mu \\ PINAW + \exp(-\eta(PICP - \mu)), PICP < \mu \end{cases} \tag{27}$$

where μ is expected value and η exponentially magnifies the difference between the PICP and μ. By minimizing the CWC function, an optimal PI is expected to be achieved.

Although the CWC$_{original}$ has been used in various interval prediction applications, deficiencies still exist [43]. The problem concerned in this paper is that the original CWC function cannot give expression of the variation of the PINAW index if the value of PICP is smaller than μ. Under this condition, the CWC function can't reasonably evaluate the prediction interval. Function diagram of $CWC_{original}$ is shown in Fig. 4, from which it can be seen that the surface of $CWC_{original}$ is uneven between the PINAW axis and the 1-PICP axis. It's obvious that the $CWC_{original}$ is more sensitive to the variation of the 1-PICP compared with PINAW. A simple case would be cogent. A simple example could be raised to illuminate this problem. Assume that the parameters of $CWC_{original}$ function is set as Table 3 and two PIs have been constructed. The PICP and PINAW of one PI are 0.89 and 0.05, while those values of the other PI are 0.9 and 0.3. The $CWC_{original}$ values are 1.212 and 0.300 respectively, which means the quality of the second PI is better. However, it's manifested that the first one should be the winner for it holds a similar coverage probability and a greatly better PI width.

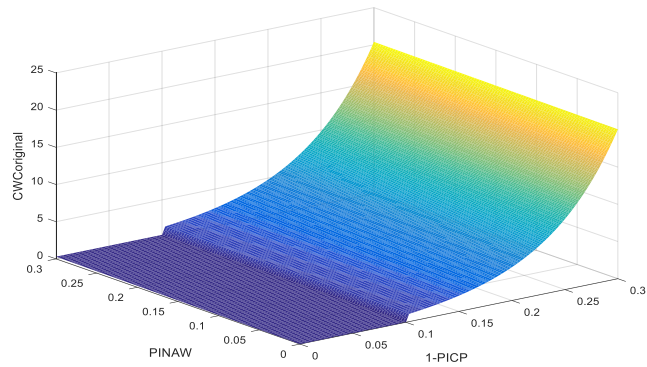

Fig. 4 Function diagram of CWC$_{orignal}$

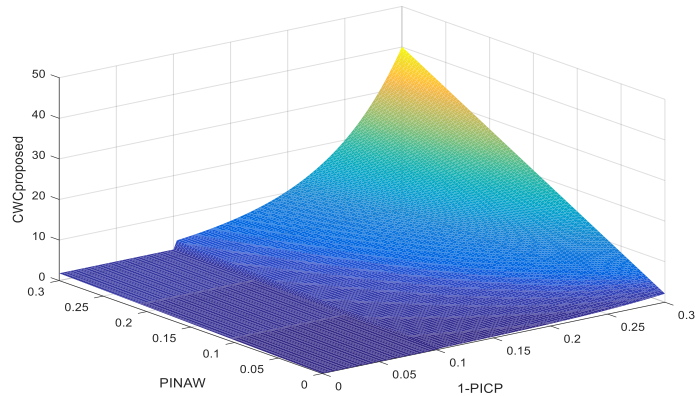

**Fig. 5** Function diagram of CWC$_{proposed}$

In order to solve above-mentioned problem, a new CWC is proposed and defined as:

$$CWC_{proposed} = \begin{cases} \beta \cdot PINAW, PICP >= \mu \\ (\alpha + \beta * PINAW) \cdot (1 + \exp(-\eta(PICP - \mu))), PICP < \mu \end{cases} \quad (28)$$

In the proposed function, the addition operation between PINAW and PICP is changed to be multiplication operation, which can solve the problem that PINAW loses control of CWC. Hyper-parameter β is applied to linearly magnify the PINAW and a small hyper-parameter α is added to avoid the problem of CWC value vanishing, once PINAW becomes zero. The parameter η exponentially magnifies the difference between the PICP and μ. Function diagram of $CWC_{proposed}$ is showed in Fig. 5, from which it can be seen that the surface of $CWC_{proposed}$ decrease uniformly between the PINAW axis and the 1-PICP axis as the dependent variables reduce.

## 4. Experiments

To evaluate the performance of the new WPIP method, the popular LUBE approaches based on SVM, KELM and ANN are compared to prove the superiority of the LSTM model. In the following numerical experiments, the effectiveness of the proposed CWC function will be evaluated at first. Then, the LUBE approaches based on SVM, KELM and ANN models, with the new CWC function as the objective function, will be compared to the new LUBE approach based on LSTM to verify the advantages of the proposed model in applications of WPIP.

To evaluate the performance of different models quantitatively, the index improvement ratio is defined as:

$$P_{index} = \frac{v_A - v_B}{v_A} \times 100\% \tag{29}$$

where $v_A$ and $v_B$ are the index values of model A and model B, respectively.

### 4.1 Data and model description

Wind power datasets are obtained from the National Renewable Energy Laboratory (NREL) website [22], which were established in 2012. Wind power data of four seasons from four farms located at two different offshore sites and two onshore sites are considered in the present work. The specific information on wind power data is exhibited in Table 1. All the wind farms are in different geographical locations and have a rated capacity of 16 MW. Data during a week with 10-min resolution includes 1008 point is used as a case for model evaluations and there are totally 16 cases in this study. The forecast horizon of LSTM and its comparison models is 10 min ahead. For each case, we use the former 5 days to train the model and the latter 2 days to test the model. In the prediction process of ANN, SVM, KELM, and LSTM model, the former 9 points are applied as input to predict the latter one point in each time step. Indices of PICP, PINRW, INAD, and CWC are calculated based on the test data.

LUBE approaches based on SVM, KELM [8] and ANN [43] are adopted to prove the superiority of the new LSTM model. Parameters of the LSTM model are selected by means of trial-and-error, while parameters of others models are set as references suggested. Other parameters of the different models used in comparative studies are presented in Table 2. PI is directly produced by ANN model, whose weights and bias are optimized by GSA [29]. The KELM based LUBE approach [8] provides a convenient way to transfer the point prediction model to the interval prediction model by equation (30). Upper bound and lower bound of PI are separately produced by KELM 1 and KELM 2. Penalty coefficient C and kernel parameter K from KELM 1 and KELM 2 as well as $\sigma_1$ and $\sigma_2$ are optimized by GSA. Since the SVM model is widely used in point prediction, the SVM based LUBE model is built in this paper for comparison by referring the KELM based approach.

$$Y_i = [y_i - \sigma_1, y_i + \sigma_2] \tag{30}$$

where $Y_i$ is the values of upper and lower bound, $y_i$ is prediction label and $\sigma$ is width factor.

Table 1

Information of 16 cases of wind power data

| No. of Datasets | Locations | Feature | Site | latitude/longitude | Seasons |
|---|---|---|---|---|---|
| 1-4 | Maine | offshore | no 114903 | 43.71859/-69.3235 | spring, summer, autumn, winter |
| 5-8 | Rhode Island | offshore | no 90994 | 40.96344/-71.4335 | spring, summer, autumn, winter |
| 9-12 | North Carolina | onshore | no 39835 | 36.4321/-76.2366 | spring, summer, autumn, winter |
| 13-16 | Virginia | onshore | no 38539 | 36.87544/-80.2524 | spring, summer, autumn, winter |

Table 2

Parameter setting of four models

| **Model** | Parameter setting |
|---|---|
| SVM | Parameters of GSA for optimization: population size is 20; maximum number of iterations is 100 |
| KELM | Parameters of GSA for optimization: population size is 20; maximum number of iterations is 100 |
| ANN | Parameters of network: number of layers is 3; number of neurons in input, hidden and output layers is 9, 10 and 2 respectively;<br>Parameters of GSA: population size is 100; maximum number of iterations is 200 |
| LSTM | Parameters of network: Dimension of $H_r$ and $C_r$ is 64; number of LSTM cell is 9; number of fully connected layers is 4; number of neurons in the 4 layers is 64, 32, 16, 8 and 2;<br>Hyper-parameters: $k_1=2$; $k_2=1$; $\lambda=4$; $\rho=0.9$; $\varepsilon=0.001$; $\delta=10^{-6}$. |

## 4.2 Comparison of CWC index

To compare the performance of the proposed CWC index, the two functions, $CWC_{original}$ and $CWC_{proposed}$, are applied as the objective function in the LUBE model based on ANN in turn. The hyper-parameters of two CWC functions are shown in Table 3. The value of α and β is set by means of trial-and-error. μ and η refer to paper [43]. Three cases of datasets extracted from different seasons in different regions are chosen to validate the effectiveness of the proposed CWC index.

Table 3

The hyper-parameters in CWC

| Parameters | Parameter | Numerical value |
|---|---|---|
| $CWC_{original}$ | μ | 0.9 |
| | η | 15 |
| $CWC_{proposed}$ | μ | 0.9 |
| | η | 15 |
| | α | 0.1 |
| | β | 6 |

Table 4

Indices of PIs obtained by $CWC_{original}$ and $CWC_{proposed}$

| Case | Objective function | Index |
|---|---|---|

|  |  | PICP | PINRW | INAD |
|---|---|---|---|---|
| Case 1 | $CWC_{original}$ | 0.8264 | 0.2220 | 0.6954 |
|  | $CWC_{proposed}$ | 0.8681 | 0.1760 | 0.6128 |
| Case 2 | $CWC_{original}$ | 0.7674 | 0.3252 | 0.2466 |
|  | $CWC_{proposed}$ | 0.9861 | 0.2205 | 0.2031 |
| Case 3 | $CWC_{original}$ | 0.9340 | 0.3196 | 0.0125 |
|  | $CWC_{proposed}$ | 0.9375 | 0.2847 | 0.0145 |

The WPIP results obtained by using the two objective functions are presented in Table 4, in which indices of PICP, PINRW, and INAD are compared to evaluate the PI quality. The results show that the PIs obtained by applying the new CWC function are markedly better than those of the original CWC function in terms of all three indices in Case 1 and Case2. As a comprehensive index, $CWC_{proposed}$ is better than $CWC_{original}$.

### 4.3 Comparison of different models

In this subsection, the four approaches are fully compared by adopting all the 16 cases of wind power datasets. To eliminate the randomness of heuristic optimization, the entire modeling of training and test are repeated 20 times and the mean values of indices are kept for comparison. As for the SVM, KELM, and ANN based LUBE approaches, the $CWC_{proposed}$ index is chosen as the objective function for model training. To evaluate performances of the models, indices of PICP, PINRW, INAD, CWC (the proposed one) and TIME (for training) are used in the following experiments.

Table 5 and Table 6 present the comparative experimental results on indices values of different models, where the offshore wind power datasets and onshore wind power datasets are adopted respectively. The average values obtained by the four models on the 16 cases are presented in Table 7. By comparing the index values obtained by different interval prediction models, it is manifested that the proposed method achieves the best performance as a whole and that the improvement over the traditional LUBE method is significant. Through a preliminary estimation, it is found that the LSTM based LUBE is effective in WPIP applications. More specific and special analyses on indices of PICP, PINRW, CWC, and TIME are conducted in following.

Table 5

The indices comparison of different models by applying offshore wind power datasets

| AVERAGE | | Offshore wind field in Maine | | | | Offshore wind field in Rhode Island | | | |
|---|---|---|---|---|---|---|---|---|---|
|  |  | spring | summer | autumn | winter | spring | summer | autumn | winter |
| SVM | PICP | 0.9688 | 0.8125 | 0.8160 | 0.9444 | 0.9479 | 0.9444 | 0.9514 | 0.9757 |
|  | PINRW | 0.1407 | 0.0791 | 0.2147 | 0.1977 | 0.3308 | 0.4045 | 0.1921 | 0.2356 |
|  | INAD | 0.0038 | 0.0706 | 0.0571 | 0.0153 | 0.0062 | 0.0018 | 0.0117 | 0.0009 |
|  | CWC | 0.8187 | 2.5853 | 5.8514 | 1.1819 | 4.3317 | 2.3893 | 1.1447 | 1.4094 |
|  | TIME | 67.277 | 44.305 | 53.854 | 65.034 | 52.532 | 66.542 | 54.339 | 51.971 |
| KELM | PICP | 0.9931 | 0.8299 | 0.7910 | 0.9410 | 0.9757 | 0.9896 | 0.9097 | 0.9583 |
|  | PINRW | 0.1852 | 0.0960 | 0.1133 | 0.2032 | 0.4412 | 0.2593 | 0.1458 | 0.1385 |
|  | INAD | 0.0005 | 0.0467 | 0.1511 | 0.0196 | 0.0018 | 0.0004 | 0.0257 | 0.0031 |
|  | CWC | 1.0984 | 2.6026 | 4.6589 | 1.2148 | 2.6431 | 1.5520 | 0.8736 | 0.8078 |
|  | TIME | 38.339 | 36.297 | 36.224 | 36.427 | 36.291 | 36.435 | 36.177 | 36.290 |
| ANN  | PICP  | 0.9153  | 0.8458  | 0.8042  | 0.9580  | 0.9210  | 0.8731  | 0.8030  | 0.9389  |
|      | PINRW | 0.2922  | 0.2150  | 0.2108  | 0.3245  | 0.1724  | 0.3375  | 0.2219  | 0.3543  |
|      | INAD  | 0.0108  | 0.0301  | 0.0611  | 0.0143  | 0.0152  | 0.0128  | 0.0558  | 0.0005  |
|      | CWC   | 2.3312  | 4.3274  | 5.3357  | 1.8710  | 0.8667  | 5.1032  | 7.4589  | 4.1988  |
|      | TIME  | 22.761  | 22.705  | 22.713  | 22.706  | 22.752  | 22.743  | 22.719  | 22.745  |
| LSTM | PICP  | 0.9792  | 0.8958  | 0.9063  | 0.9549  | 0.9826  | 0.9896  | 0.9097  | 1.0000  |
|      | PINRW | 0.1493  | 0.1970  | 0.2710  | 0.1917  | 0.1631  | 0.2114  | 0.1639  | 0.2266  |
|      | INAD  | 0.0017  | 0.0175  | 0.0208  | 0.0142  | 0.0011  | 0.0018  | 0.0245  | 0.0000  |
|      | CWC   | 0.8105  | 2.1656  | 1.5474  | 0.8853  | 0.8553  | 1.0941  | 0.8838  | 1.1297  |
|      | TIME  | 11.422  | 11.346  | 11.348  | 11.374  | 11.327  | 11.359  | 11.349  | 11.348  |

Table 6

The indices comparison of different models by applying onshore wind power datasets

| AVERAGE |       | Onshore wind field in North Carolina |        |        |        | Onshore wind field in Virginia |        |        |        |
|---------|-------|--------|--------|--------|--------|--------|--------|--------|--------|
|         |       | spring | summer | autumn | winter | spring | summer | autumn | winter |
| SVM     | PICP  | 0.8194 | 0.9271 | 0.9340 | 0.9028 | 0.9410 | 0.8153 | 0.9757 | 0.8958 |
|         | PINRW | 0.1490 | 0.2016 | 0.0958 | 0.1439 | 0.2642 | 0.2687 | 0.1836 | 0.1626 |
|         | INAD  | 0.0849 | 0.0101 | 0.0164 | 0.0220 | 0.0172 | 0.0931 | 0.0042 | 0.0229 |
|         | CWC   | 4.2761 | 1.1450 | 0.5627 | 0.8116 | 1.4450 | 7.8419 | 0.9894 | 2.1763 |
|         | TIME  | 49.316 | 47.076 | 57.185 | 81.396 | 107.764| 44.363 | 59.924 | 72.331 |
| KELM    | PICP  | 0.8854 | 0.8819 | 0.8750 | 0.8715 | 0.9167 | 0.8160 | 0.9861 | 0.9479 |
|         | PINRW | 0.1441 | 0.1969 | 0.1030 | 0.1170 | 0.1241 | 0.1320 | 0.1558 | 0.2277 |
|         | INAD  | 0.0473 | 0.0255 | 0.0402 | 0.0340 | 0.0435 | 0.0668 | 0.0035 | 0.0061 |
|         | CWC   | 2.1349 | 2.9576 | 1.7497 | 2.0188 | 0.7345 | 3.9813 | 0.9019 | 1.3655 |
|         | TIME  | 36.614 | 36.796 | 36.824 | 36.895 | 37.050 | 36.952 | 37.177 | 36.985 |
| ANN     | PICP  | 0.8528 | 0.8710 | 0.9542 | 0.9031 | 0.8806 | 0.8722 | 0.8639 | 0.9226 |
|         | PINRW | 0.2512 | 0.2894 | 0.3204 | 0.3116 | 0.2627 | 0.2651 | 0.2269 | 0.3112 |
|         | INAD  | 0.0382 | 0.0258 | 0.0111 | 0.0141 | 0.0241 | 0.0345 | 0.0106 | 0.0144 |
|         | CWC   | 3.665  | 5.218  | 1.927  | 2.907  | 4.027  | 3.401  | 6.877  | 3.217  |
|         | TIME  | 23.124 | 23.100 | 23.213 | 23.209 | 23.523 | 23.409 | 23.414 | 23.393 |
| LSTM    | PICP  | 0.9063 | 0.9410 | 0.9653 | 0.9549 | 0.9722 | 0.9028 | 0.9479 | 0.9340 |
|         | PINRW | 0.2019 | 0.1865 | 0.1724 | 0.1717 | 0.1256 | 0.1599 | 0.1287 | 0.1607 |
|         | INAD  | 0.0267 | 0.0107 | 0.0059 | 0.0066 | 0.0168 | 0.0309 | 0.0053 | 0.0130 |
|         | CWC   | 1.0976 | 1.0152 | 0.9839 | 0.9079 | 0.6130 | 0.9049 | 0.6883 | 0.8704 |
|         | TIME  | 11.329 | 11.320 | 11.372 | 11.374 | 11.367 | 11.360 | 12.371 | 12.639 |

Table 7

Comparison of average values of different models achieved on the 16 cases datasets

| Model/index | PICP | PINRW | INAD | CWC  | TIME  |
|-------------|------|-------|------|------|-------|
| SVM         | 0.91 | 0.20  | 0.03 | 2.44 | 60.95 |
| KELM        | 0.91 | 0.17  | 0.03 | 1.96 | 36.74 |
| ANN         | 0.89 | 0.27  | 0.02 | 3.92 | 23.01 |
| LSTM | 0.95 | 0.18 | 0.01 | 1.03 | 11.50 |

For more intuitive comparison, histograms of the two key indices, the PICP and PINRW, are presented contrastively in Fig. 6, where the four models are compared in all cases. In this figure, a small PINRW and large PICP are desirable. From Fig. 6, it's found that the LSTM model wins in the PICP index on 11 cases, while in rest of cases results of the LSTM take the second place and are just slightly lower than the best values. From Table7, it is found that the average value of PICP index is 0.95, which is significantly larger than those of other models. It means a desirable coverage probability of PI could be guaranteed. As for the PINRW index, the average values of the four models on the 16 cases are 0.2, 0.17, 0.27 and 0.18 respectively, as stated in Table 7. Although the LSTM model is a little under shadowed by the KELM model in this regard, the index values are still excellent and stable. Considering that the corresponding PICP of the KELM model is only 0.91, the overall performance of the LSTM model is more satisfactory. In general, the proposed model reaches a good balance between the indices of PINRW and PICP.

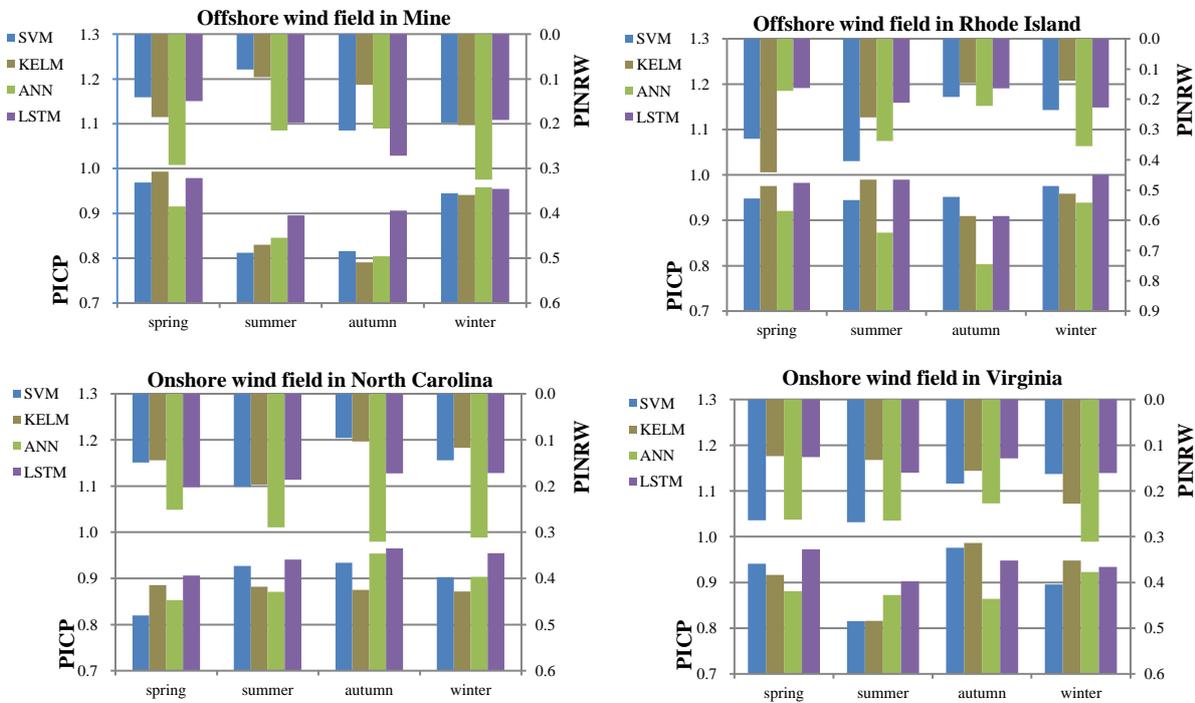

**Fig. 6** Comparison of PICP and PINRW

For evaluating the effectiveness and efficiency, the CWC index and TIME index are compared and analyzed in Fig. 7 and Fig. 8. Clearly, from Fig. 7, it can be observed that LSTM achieves the smallest values of CWC index in 12 cases, and shows a slightly worse performance in the rest cases with competitive values. From Table7, it is found that the average value of CWC index is 1.03, which is significantly better than those of other models. As introduced above, the CWC index is essential for PI evaluation in interval prediction applications. By comparing the CWC index, a conclusion can be drawn that the PIs obtained by the LSTM model is with the best quality in the comparative experiments.

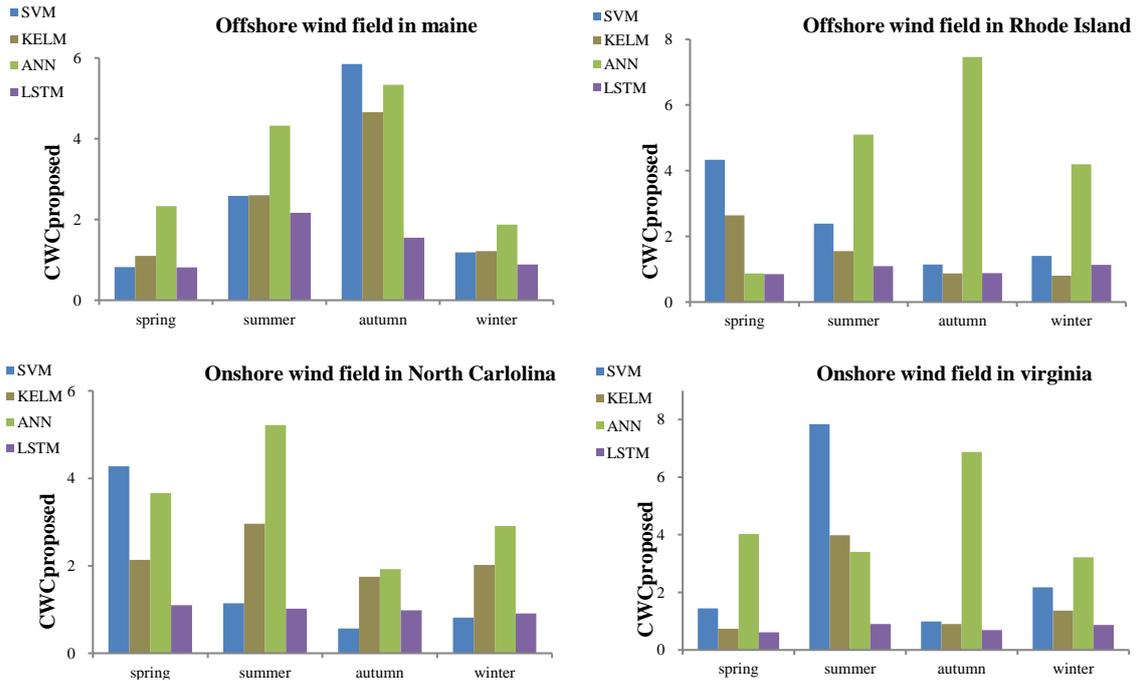

**Fig. 7** Comparison of CWC

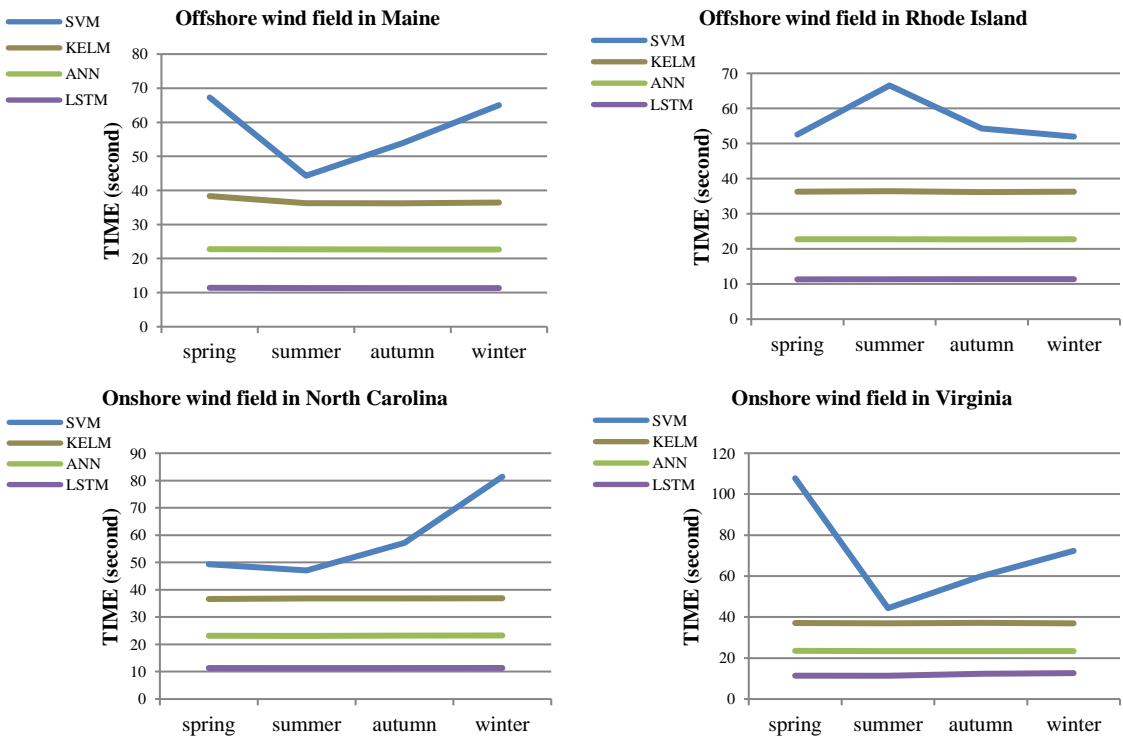

**Fig. 8** Comparison of TIME

From Fig. 8, it is manifested that the TIME index obtained by the LSTM model is dramatically reduced compared with those of SVM, KELM, and ANN in all cases. Compared to the most time-consuming model, namely the SVM model, with around 60 seconds of training time, it only takes nearly 10 seconds to train the LSTM model. Compared with other models, training time of the LSTM model is more stable. Obviously, the training process of the LSTM model is more efficient in comparison to those traditional LUBE approaches.

To evaluate promotion of LSTM model quantitatively, the promotion ratio of the LSTM model over other models on indices of CWC and TIME are exhibited in Table 8 and Table 9. From these tables, it is found that most of the increasing ratios of CWC are around 50% in all 48 groups of comparison, except for negative results in 4 groups. However, it is worth noticing that the average increasing ratio of CWC index obtained by the LSTM over other models is 45%. The reduction ratio of TIME is at least more than 40% in all cases and the average reduction ratio is 66%. From these statistic results, it is easy to find out that the LSTM model makes a notable improvement in terms of both effectiveness and efficiency.

Table 8

Improved percentages compared with comparison model (offshore)

| Extant models vs. LSTM | | Offshore wind field in Maine | | | | Offshore wind field in Rhode Island | | | |
|---|---|---|---|---|---|---|---|---|---|
| | | spring | summer | autumn | winter | spring | summer | autumn | winter |
| SVM | CWC | 1.01% | 16.24% | 73.55% | 25.10% | 80.25% | 54.21% | 22.80% | 19.84% |
| | TIME | 83.02% | 74.39% | 78.93% | 82.51% | 78.44% | 82.93% | 79.11% | 78.17% |
| KELM | CWC | 26.21% | 16.79% | 66.79% | 27.12% | 67.64% | 29.50% | -1.17% | -39.85% |
| | TIME | 70.21% | 68.74% | 68.67% | 68.78% | 68.79% | 68.82% | 68.63% | 68.73% |
| ANN | CWC | 65.23% | 49.96% | 71.00% | 52.69% | 1.31% | 78.56% | 88.15% | 73.09% |
| | TIME | 49.82% | 50.03% | 50.04% | 49.91% | 50.22% | 50.05% | 50.05% | 50.11% |

Table 9

Improved percentages compared with comparison model (onshore)

| Extant models vs. LSTM | | Onshore wind field in North Carolina | | | | Onshore wind field in Virginia | | | |
|---|---|---|---|---|---|---|---|---|---|
| | | spring | summer | spring | summer | spring | summer | spring | summer |
| SVM | CWC | 74.33% | 11.33% | -74.88% | -11.87% | 57.58% | 88.46% | 30.43% | 60.01% |
| | TIME | 77.03% | 75.95% | 80.11% | 86.03% | 89.45% | 74.39% | 79.36% | 82.53% |
| KELM | CWC | 48.59% | 65.67% | 43.77% | 55.03% | 16.54% | 77.27% | 23.69% | 36.26% |
| | TIME | 69.06% | 69.23% | 69.12% | 69.17% | 69.32% | 69.26% | 66.72% | 65.83% |
| ANN | CWC | 70.05% | 80.54% | 48.93% | 68.77% | 84.78% | 73.39% | 89.99% | 72.95% |
| | TIME | 51.01% | 50.99% | 51.01% | 50.99% | 51.68% | 51.47% | 47.16% | 45.97% |

For further validation of the superiority of the new model proposed in this paper, two cases (case 1 and case 12), one offshore case and one onshore case, are chosen to exhibit PI quality by the graphical presentation, while the four models are compared. To carry out a fair comparison, for each model, prediction results of one run in the 20 runs on a dataset, whose CWC value is closest to the average value of the 20 runs, are presented. Comparison of the predicted intervals for offshore case is shown in Fig. 9, and that of onshore case is shown in Fig. 10.

From Fig.9, it is manifested that all models perform excellent in terms of coverage probabilities. The LSTM model outperforms the SVM model, KELM model and the ANN model with narrow PI widths. In Fig 10, it seems that the PI widths obtained by the SVM model and KELM model are wonderful. However, the coverage probability is not good, while it's observed that the real values fall outside the intervals in the marked zooms. Therefore, it is clear that the LSTM model achieves the best performance in producing high quality PIs in WPIP applications.

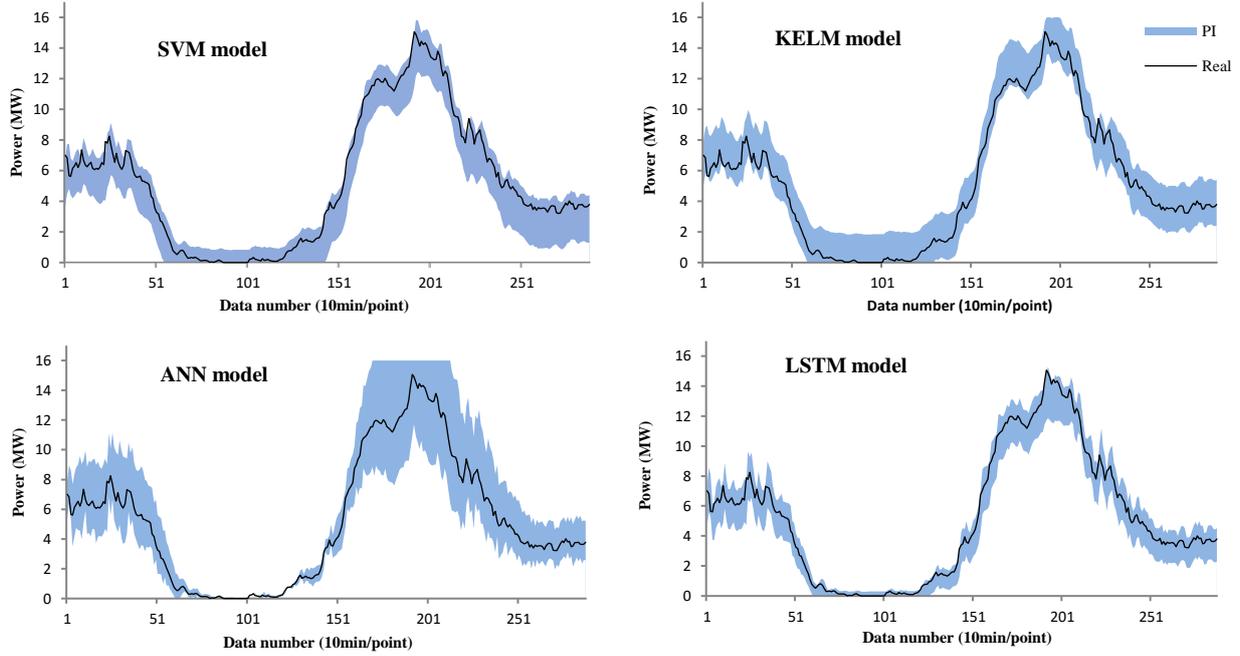

**Fig. 9** Comparison of the predicted intervals for WPIP in Maine offshore wind farm in spring

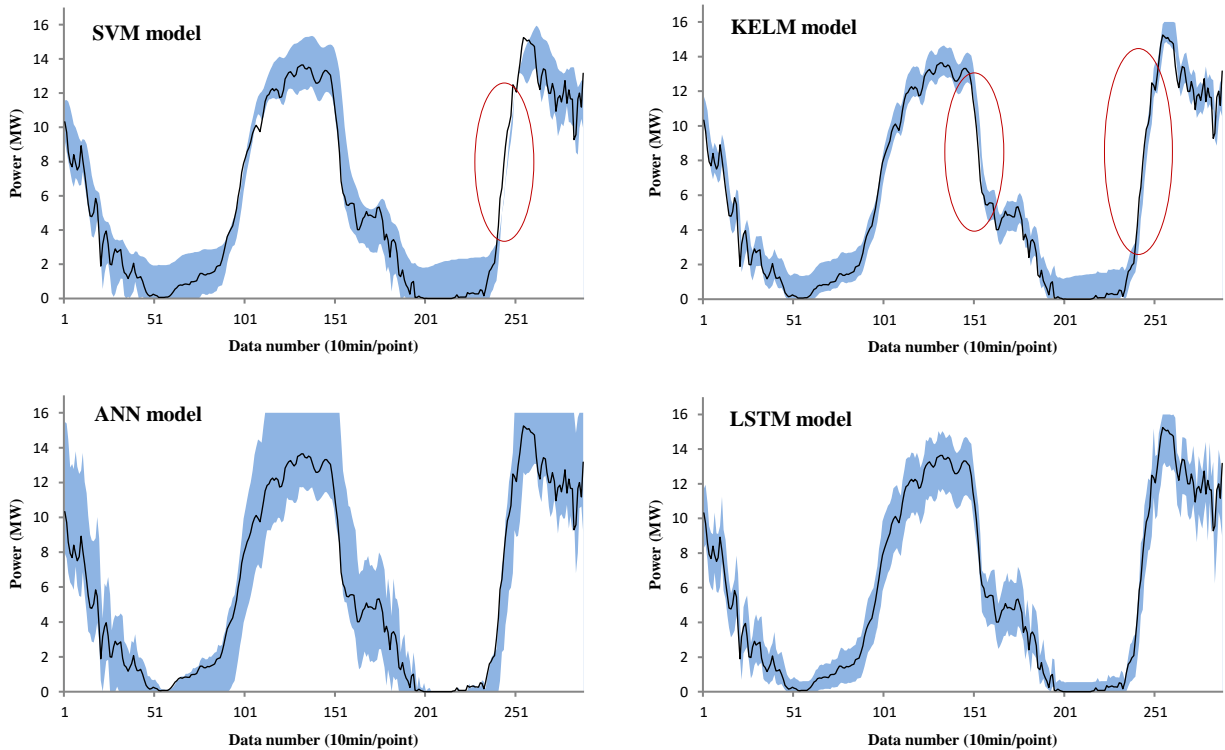

**Fig. 10** Comparison of the predicted intervals for WPIP in North Carolina onshore wind farm in winter

### 4.4 Mass data validation

We selected wind power data from the California site in 2011 and 2012. A season's data at a 10-min resolution consisted of 17,280 points were used as a training dataset for model evaluations. The forecast horizon of LSTM and its comparison models is 10 min ahead. There were four cases in this study. For each case, we use the former 3 months to train the model and the latter month to test the model. For example, the "spring case" use data from January to March to train and April data to test.

To further validate the effectiveness of the new LSTM model, statistical methods, bootstrap method [2][42] and ARIMA method [34][24] relied on large-scale data were employed for comparison. Confidence level of bootstrap method and ARIMA method are set as 0.92. Moreover, ANN model in this article has been included in the comparison. The average values of PICP, PINRW, and CWC are summarized in Table 10 and the corresponding bar graphs in Fig. 11.

From Fig. 11, it is manifested that the LSTM model outperforms its competitor models with high coverage probabilities and narrow PI widths. Although the PICP obtained by the LSTM model in winter dataset are inferior than its comparison model, the difference of their value, 0.9367, 0.9242, 0.9293 and 0.9240 (LSTM) in Table 10, is small. According to CWC, LSTM model remains smallest in four datasets. Therefore, it is clear that the LSTM model achieves the best performance in producing high quality PIs in mass datasets.

Table 10

Results of mass data forecasting on California datasets

| **Average value of 10 times** | | spring | summer | autumn | winter |
|---|---|---|---|---|---|
| **ANN** | PICP | 0.9488 | 0.9587 | 0.9562 | 0.9367 |
|  | PINRW | 0.2249 | 0.1951 | 0.2573 | 0.2244 |
|  | CWC | 1.3495 | 1.1704 | 1.5436 | 1.3461 |
| **ARIMA** | PICP | 0.9227 | 0.9275 | 0.9253 | 0.9242 |
|  | PINRW | 0.0956 | 0.0905 | 0.0770 | 0.0875 |
|  | CWC | 0.5738 | 0.5428 | 0.4617 | 0.5248 |
| **Bootstrap** | PICP | 0.9263 | 0.9367 | 0.9240 | 0.9293 |
|  | PINRW | 0.0924 | 0.0822 | 0.0697 | 0.0991 |
|  | CWC | 0.5546 | 0.4933 | 0.4179 | 0.5947 |
| **LSTM** | PICP | 0.9760 | 0.9545 | 0.9788 | 0.9240 |
|  | PINRW | 0.0396 | 0.0100 | 0.0334 | 0.0197 |
|  | CWC | 0.2377 | 0.0987 | 0.2280 | 0.4101 |

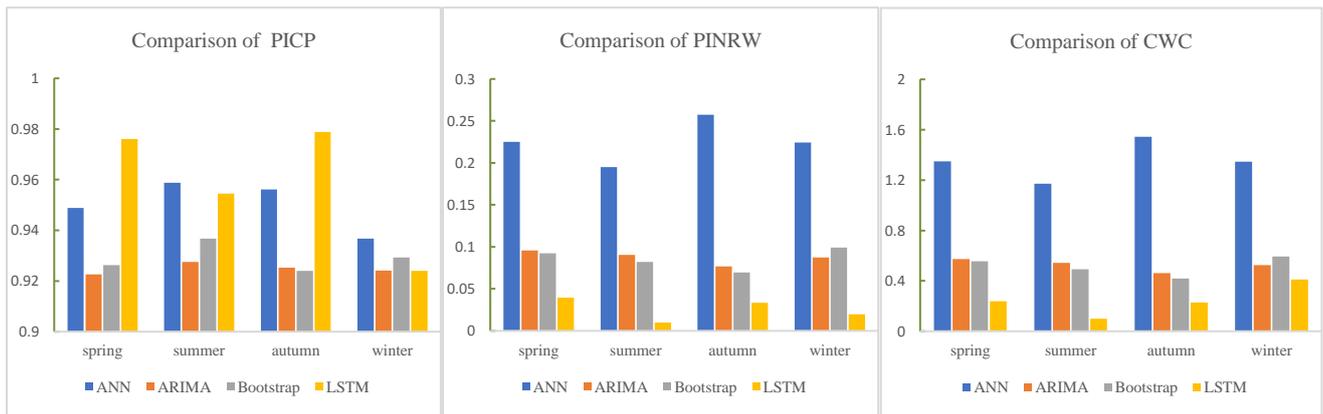

**Fig. 11** Comparison in terms of PICP, PINRW, and CWC using mass data

## 4.5 Discussion on applications

Wind power forecasting is significant for wind power generation and management, which provides fundamental support for wind turbine control and scheduling of hybrid power system containing wind power. It is widely known that accurate and stable wind power forecasting plays a vital role in wind turbine power generation. Compared with point prediction, wind power interval prediction can quantify the range of changes in prediction results due to uncertain factors at a set of confidence levels that determine the predicted interval at the observed value, and can provide a comprehensive reference to support power system planning.

The detailed roles of wind power interval prediction for design and operation of a power system including wind farms are summarized as follows:

Firstly, WPIP models can provide sufficient information for decision-makers who are making plans for wind turbine power generation. Then decision-makers can make a detailed schedule for adjusting wind turbines to ensure the maximum yield of wind energy. Secondly, the balance of power supply and demand is essential, which plays a fundamental role in sustainable energy management and economically efficient operation. Overload, on one hand, will result in an increase in start-up and long-term costs due to the inherent difficulties in storing electricity while under-load, on the other hand, will negatively affect the quality of power supply, rendering it incapable of satisfying regular power demands and potentially compromising the safety and stability of the power system [37].

## 5. Conclusion

In this paper, a new interval prediction model based on long short-term memory network and lower upper bound estimation method is proposed for wind power interval prediction. To the best of our knowledge, it is the first time that a long short-term memory model has been built in the frame of lower upper bound estimation for interval prediction. A new model composed of three parts, namely the long short-term memory model, the fully connected layers and the rank ordered model, is designed, while two objective functions are proposed for implementing the training method based on the root mean square back propagation algorithm. Three popular lower upper bound estimation methods based on kernel extreme learning machine, support vector machine and artificial neural network are compared to prove the superiority of the long short-term memory model. An index of improved coverage width criterion is proposed to evaluate those intercomparable interval prediction models effectively.

According to the experimental results, prediction interval coverage probability, normalized average deviation, coverage width criterion and training time obtained by the models based on long short-term memory are dramatically better than methods based on kernel extreme learning machine, support vector machine and artificial neural network with average 45% promotion on the index of coverage width criterion. Besides, the model based on long short-term memory takes great advantage in terms of efficiency with reduction of average 66% of time consumption.

Furthermore, mass data experiments added to test the superiority of long short-term memory model in mass data. The results show that long short-term memory model achieves significant better performance in producing high quality prediction interval than traditional model in mass datasets.

## Acknowledgements

This paper is supported by the National Natural Science Foundation of China (No. 51879111, No. 51679095) and the Applied Fundamental Frontier Project of Wuhan Science and Technology Bureau (No. 201801040101269).